\documentclass{jetpl}
\twocolumn

\usepackage{amsmath,amssymb,epsfig,color,pstricks,bm,graphics,alltt}

\usepackage[colorlinks,plainpages=false,linkcolor=black,urlcolor=blue,citecolor=black,pdfpagemode=UseNone,pdfstartview=FitBH]{hyperref}

\definecolor{MyDarkBlue}{rgb}{0,  0.3,  0.9}
\definecolor{MyDarkBlack}{rgb}{0,  0,  0}

\begin{document}

%%% article in English
\lat

%%% article title
\title{Electronic Structure, Topological Phase Transitions and Superconductivity 
in (K,Cs)$_x$Fe$_2$Se$_2$}

%%% article title - for colontitle (at the top of the page)
\rtitle{Fermi surface topology of (K,Cs)$_x$Fe$_2$Se$_2$ superconductors}

%%% article title - for table of contents (usualy identical with \title)
\sodtitle{Electronic Structure, Topological Phase Transitions and 
Superconductivity in (K,Cs)$_x$Fe$_2$Se$_2$}

%%% author(s) ( + e-mail)
\author{I.\ A.\ Nekrasov\thanks{E-mail: nekrasov@iep.uran.ru}, M.\ V.\ Sadovskii\thanks{E-mail: sadovski@iep.uran.ru}}

%%% author(s) - for colontitle (at the top of the page)
\rauthor{I.\ A.\ Nekrasov, M.\ V.\ Sadovskii}

%%% author(s) - for table of contents
\sodauthor{Nekrasov, Sadovskii }

%%% author(s) - for table of contents
\sodauthor{Nekrasov, Sadovskii}

%%% author's address(es)
\address{Institute for Electrophysics, Russian Academy of Sciences, 
Ural Branch, Amundsen str. 106,  Ekaterinburg, 620016, Russia}

%%% dates of submition & resubmition (if submitted once, second argument is *)
\dates{December 2010}{*}

%%% abstract
\abstract{
We present LDA band structure of novel hole doped
high temperature superconductors (T$_c\sim$ 30K)~K$_x$Fe$_2$Se$_2$ and 
Cs$_x$Fe$_2$Se$_2$ and compare it with previously studied electronic structure 
of isostructural FeAs superconductor BaFe$_2$As$_2$ (Ba122). 
We show that stoichiometric KFe$_2$Se$_2$ and CsFe$_2$Se$_2$ have rather 
different Fermi surfaces as compared with Ba122. However at about
60\% of hole doping Fermi surfaces of novel materials closely resemble those
of Ba122. In between these dopings we observe a number of topological Fermi 
surface transitions near the $\Gamma$ point in the Brillouin zone.
Superconducting transition temperature T$_c$ of new systems is apparently
governed by the value of the total density of states (DOS) at the Fermi level.
}


\maketitle

The FeAs based high-temperature superconductors \cite{kamihara_08}
attracted a lot of attention and huge number of experimental and theoretical 
investigations have been done (for review see~\cite{UFN_90,Hoso_09})
and many are still going on. Pretty high values of superconducting transition
temperature were discovered also in Fe chalcogenides FeSe$_x$ and 
FeSe$_{1-x}$Te$_x$ \cite{FeSe}.

Structurally FeSe systems are similar to FeAs compounds and consist of layers 
of FeSe$_4$ tetrahedra. Recent discovery of intercalated K$_x$Fe$_2$Se$_2$
and Cs$_x$Fe$_2$Se$_2$ compounds produced much higher values of 
T$_c$=31K~\cite{Guo10} and 27K~\cite{Krzton10} similar to those in FeAs 122
systems \cite{UFN_90,Hoso_09}. This was followed by $T_c=$31K in 
(Tl,K)Fe$_x$Se$_2$ \cite{Fang}.

Electronic structure of Fe(S,Se,Te) materials was described in details
in Ref.~\cite{SinghFeSe}. However K$_x$Fe$_2$Se$_2$
and Cs$_x$Fe$_2$Se$_2$ have different crystal structure and are actually
isostructural to BaFe$_2$As$_2$ (Ba122). Its electronic structure was reported 
in Refs.~\cite{Nekr2,Shein, Krell}. First calculations of electronic spectrum of
K$_x$Fe$_2$Se$_2$ were described in a recent preprint~\cite{Shein_kfese}.

In this work we present comparative study of electronic structure, densities of 
states for Ba122 and K$_x$Fe$_2$Se$_2$, Cs$_x$Fe$_2$Se$_2$ systems, 
demonstrating changes of Fermi surface topology upon doping and making some 
simple estimates of superconducting $T_c$.

The K$_x$Fe$_2$Se$_2$ and Cs$_x$Fe$_2$Se$_2$ systems are isostructural
to Ba122 (for the last one see Ref.~\cite{Nekr2}) 
with ideal body centered tetragonal space group I4/mmm.
The K$_x$Fe$_2$Se$_2$ has $a$=3.9136\AA~ and $c$=14.0367\AA~
with K ions occupying $2a$, Fe -- $4d$ and Se -- $4e$ positions
with $z_{Se}$=0.3539 \cite{Guo10}.
In case of Cs$_x$Fe$_2$Se$_2$ lattice parameters are
$a$=3.9601\AA~ and $c$=15.2846\AA~and  $z_{Se}$ is 0.3439 \cite{Krzton10}.
For given crystal structures we performed 
band structure calculations within the linearized muffin-tin orbitals 
method (LMTO)~\cite{LMTO} using default settings.

In Fig.~1 we compare Ba122 band structure and different densities of states
of Ref.~\cite{Nekr2} (left) and those for K$_x$Fe$_2$Se$_2$ (black lines)
and Cs$_x$Fe$_2$Se$_2$ (gray lines) (right) for stoichiometric case of x=0.
In a bird eye view  K$_x$Fe$_2$Se$_2$ and Cs$_x$Fe$_2$Se$_2$ have nearly the same
band dispersions which to some extent are similar to those in Ba122.
However, there are some quantitative differences. 
First of all Fe-3d and Se-4p states in new systems are separated in energy
in contrast to Fe-3d and As-4p Ba122. Also Se-4p states are of about 0.7 eV 
lower than As-4p states.

\begin{figure}[h]
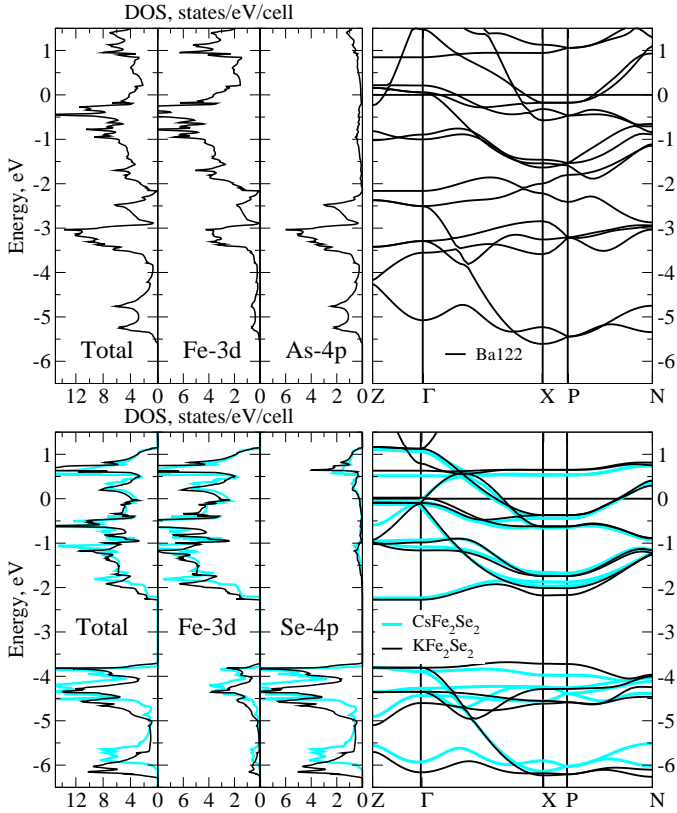

\includegraphics[clip=true,width=0.5\textwidth]{Ba122_DOS_bands.eps}
\includegraphics[clip=true,width=0.5\textwidth]{KCs_DOS_bands.eps}
\caption{Fig. 1. LDA calculated band dispersions and densities of states of 
Ba122 (upper panel) and KFe$_2$Se$_2$ (black lines) and CsFe$_2$Se$_2$ 
(gray lines)(lower panel). The Fermi level $E_F$ is at zero energy.} 
\end{figure}

Similar to Ba122 the Fermi level $E_F$ in K and Cs chalcogenides is crossed by
Fe-3d states. Detailed band structure near the Fermi level, which is
decisive for the formation of superconducting state, for the new systems
is compared with that of Ba122 in Fig. 2. To some extent Ba122 bands near $E_F$ (upper part of Fig.~2) would match
those for (K,Cs)Fe$_2$Se$_2$ if we shift them down in energy by 
about 0.2 eV. Main difference between old and new systems is seen around 
$\Gamma$ point. For (K,Cs)Fe$_2$Se$_2$ systems antibonding part Se-4p$_z$ band in the 
Z-$\Gamma$ direction forms electron-like pocket. In Ba122 corresponding band 
lies about 0.4eV higher and goes much steeper, thus it is quite far away from 
$\Gamma$ point. However, if we dope (K,Cs)Fe$_2$Se$_2$ systems 
(in a rigid band manner) with holes (as shown by horizontal lines in Fig.~2 on lower panel) 
we obtain bands around $\Gamma$ point (close to the Fermi level) very similar 
to those in case of Ba122. Namely at 60\% hole doping we obtain three hole-like 
cylinders while stoichiometric KFe$_2$Se$_2$ has one small electron pocket and 
larger hole like one and CsFe$_2$Se$_2$ has just one electron pocket near 
$\Gamma$ point. Thus, in fact under hole doping we observe several topological 
transitions of the Fermi surfaces which we shall briefly discuss below.

\begin{figure}[h]
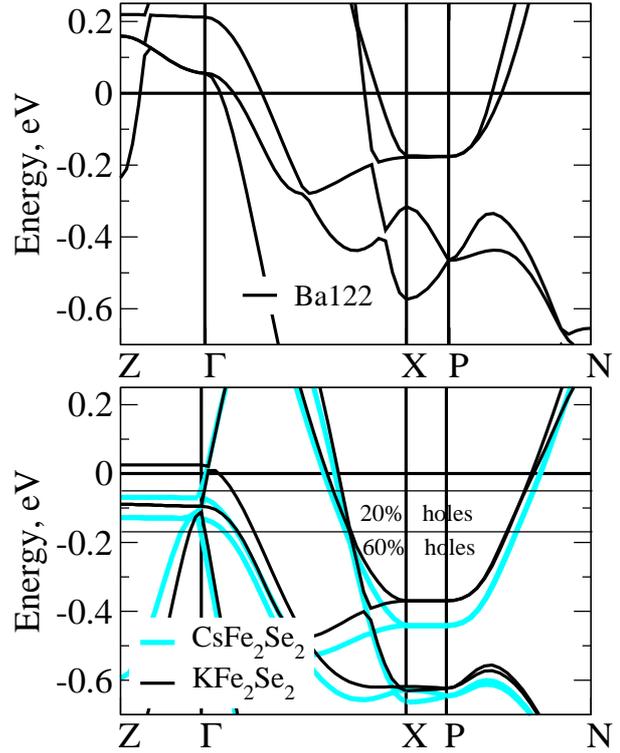

\includegraphics[clip=true,width=0.45\textwidth]{Ba122_Ef_bands.eps}
\includegraphics[clip=true,width=0.45\textwidth]{KCs_Ef_bands.eps}
\caption{Fig. 2. Top panel -- LDA calculated band dispersions in the vicinity of the Fermi 
level for Ba122; Bottom panel -- K$_x$Fe$_2$Se$_2$ (black lines) and Cs$_x$Fe$-2$Se$_2$ (gray lines).
The Fermi level is at zero energy.
Additional horizontal lines correspond to Fermi level position
for the case of 20\% and 60\% hole doping.} 
\end{figure}

To trace orbital composition of bands if Fig.~3 we show
orbital resolved densities of states for KFe$_2$Se$_2$.
Again as for Ba122 \cite{Nekr2} and other iron pnictides
mainly $t_{2g}$ states ($xy$, $xz$ and $yz$) contribute to the density of
states at the Fermi level.
The $e_g$ states ($3z^2-r^2$ and $x^2-y^2$) are almost absent in the
density of states at $E_F$.

\begin{figure}[h]
\includegraphics[clip=true,width=0.45\textwidth]{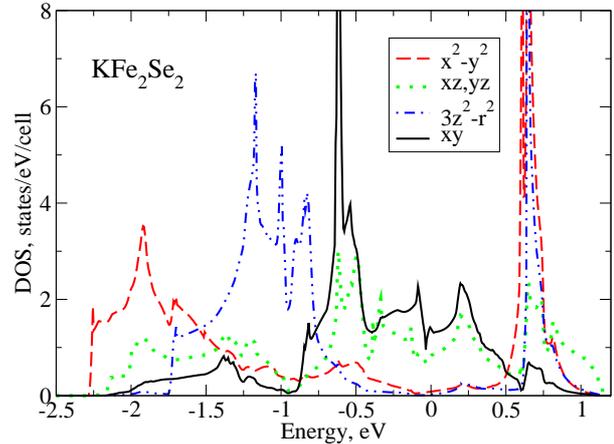}
\caption{Fig. 3. LDA calculated orbital resolved densities of states of 
KFe$_2$Se$_2$. The Fermi level is at zero energy.} 
\end{figure}

In Fig.~4 we present LDA calculated Fermi surfaces (FS) for both K (upper row)
and Cs (lower row) (K,Cs)Fe$_2$Se$_2$ compounds for different hole doping levels
of x=0 (left), x=0.2 (middle) and x=0.6 (right).
All Fermi surfaces have two almost two dimensional electron-like sheets in
the corners of the Brillouin zone with topology independent of doping level.
Compared to Ba122 FeAs system the sharp difference in the Fermi surface topology
around the center of the Brillouin zone ($\Gamma$-point) is observed at
x=0 and x=0.2. In fact, KFe$_2$Se$_2$ compound has one electron and one 
hole torus-like FS sheets while CsFe$_2$Se$_2$ has just one electron-like 
hourglass FS sheet. With hole doping KFe$_2$Se$_2$ torus transforms to 
electron-like  hourglass and hole cylinder. For 20\% hole doped Cs compound we 
get similar picture with smaller volume FS sheets of the same topology.
For x=0.6 both K and Cs new FeSe materials have Fermi surfaces quite similar to 
those in Ba122 iron pnictide (see Ref.~\cite{Nekr2}), with rather typical
hole-lke FS cylinders in the center of the Brillouin zone.

In the Ref.~\cite{Krzton10} it was shown that K and Cs compounds
follow the tendency of T$_c$ dependence on anion height in FeSe plane observed
in Ref.~\cite{hPn2}, which was plausibly explained in Ref.~\cite{Kucinskii10}
in terms of total density of states change at the Fermi level.  
Similar observation was made for related compounds SrPt$_2$As$_2$,
BaNi$_2$As$_2$ and SrNi$_2$As$_2$ in Ref.~\cite{SrPtAs}.

Now we also can make some simple BCS-like estimates of $T_c$. 
Taking the LDA calculated value of total DOS at the Fermi level N($E_F$)
3.94~states/eV/cell for K$_{x=0}$Fe$_2$Se$_2$ and 3.6~states/eV/cell for  
Cs$_{x=0}$Fe$_2$Se$_2$, $\omega_D$=350K and coupling constant 
$g$=0.21eV estimated for Ba122 (as described in Ref.~\cite{Kucinskii10}), then 
using the BCS expression for $T_c=1.14\omega_D e^{-2/gN(E_F)}$ we 
immediately obtain T$_c$=34K and 28.6K for K and Cs systems respectively
($T_c$ ratio 1.18). That is very close to experimental T$_c$ values 
31K\cite{Guo10} and 27K ($T_c$ ratio 1.15) \cite{Krzton10}.
If we take into account the fact that upon hole doping N($E_F$) grows
for both compounds up to 4.9~states/eV/cell in K and 4.7~states/eV/cell in Cs
at 60\% doping level
superconducting transition temperatures can be estimated in a similar way to
give $T_c=$57K for K system and $T_c=$52K for Cs system, showing the potential
role of doping. Thus, in accordance with our previous work on pnictides 
\cite{Kucinskii10}, the values of  T$_c$ apparently well correlate with the 
total DOS value at the Fermi level N($E_F$). It should be stressed that these
estimates do not necessarily imply electron-phonon pairing, as $\omega_D$ may
just denote the average frequency of any other possible Boson responsible for
pairing interaction (e.g. spin fluctuations). At the same time the lower values 
of $T_c$ in Cs compound in comparison to K system can be probably attributed to
the usual isotope effect.

To conclude, we investigated the band structure and Fermi surface
topology of recently discovered chalgogenide iron superconductors
K$_x$Fe$_2$Se$_2$  and Cs$_x$Fe$_2$Se$_2$ isostructural to Ba122 iron pnictide 
system at different hole doping levels.
We show that at about 60\% hole doping level both K$_x$Fe$_2$Se$_2$  and 
Cs$_x$Fe$_2$Se$_2$ energy bands and Fermi surface topologies resemble very much 
those of Ba122 FeAs system. However, at intermediate dopings there are several 
topological transitions of the Fermi surfaces with changing of number of  
(electron-like and hole-like) sheets. Also we demonstrated that T$_c$ values 
in new superconductors are well correlated with total DOS value at the Fermi 
level N($E_F$), which is related to anion height relative to Fe square lattice,
similar to that in other FeAs and Fe(Se,Te) systems.

This work is partly supported by RFBR grant 11-02-00147 and was performed
within the framework of programs of fundamental research of the Russian 
Academy of Sciences (RAS) ``Quantum physics of condensed matter'' 
(09-$\Pi$-2-1009) and of the Physics Division of RAS  ``Strongly correlated 
electrons in solid states'' (09-T-2-1011).

\onecolumn
\begin{figure}[h]
\includegraphics[clip=true,width=1\textwidth]{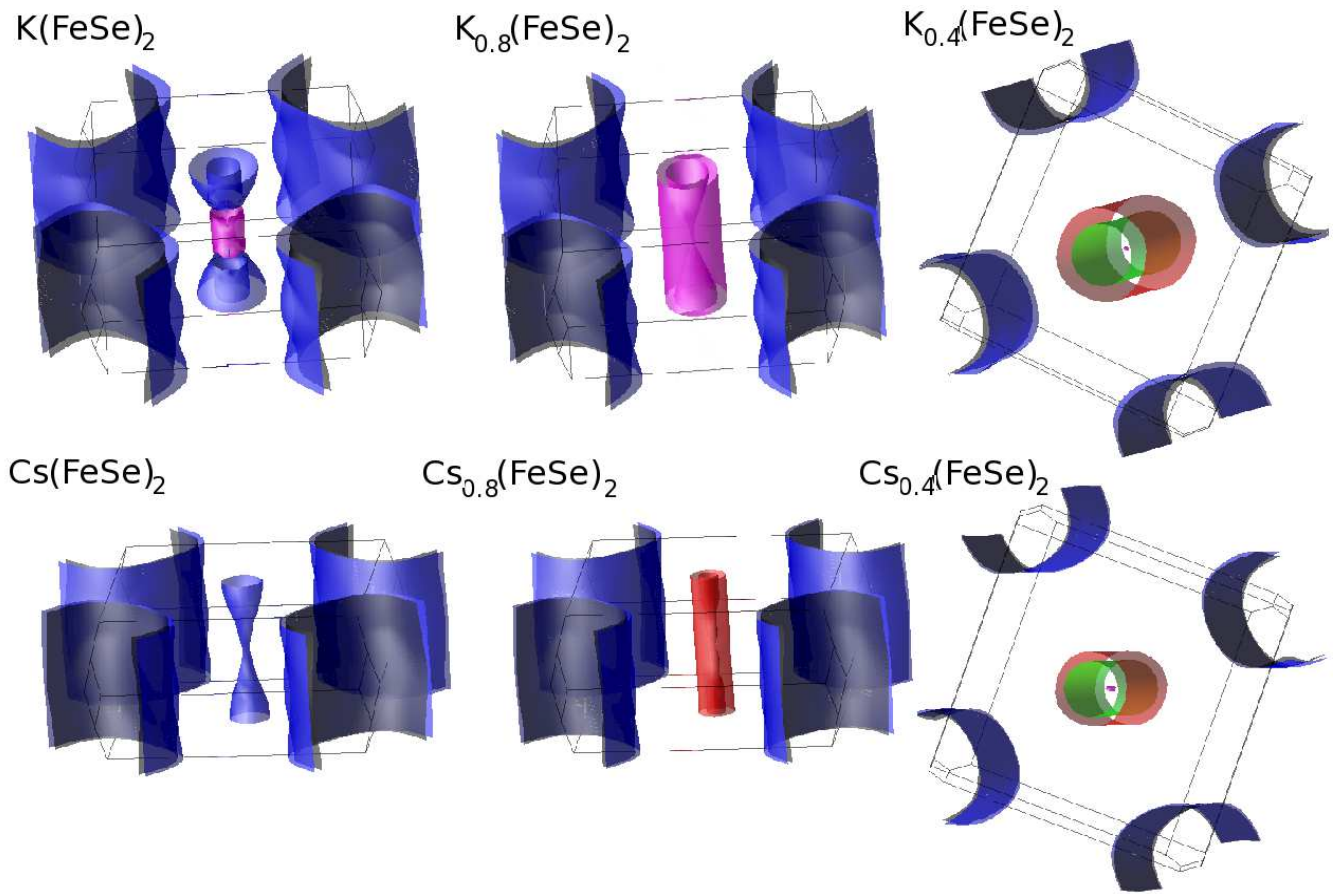}
\caption{Fig. 4. LDA calculated Fermi surfaces of
K$_x$Fe$_2$Se$_2$ (upper row) and Cs$_x$Fe$_2$Se$_2$ (lower row)
for different doping levels: x=0 -- left column, x=0.2 -- middle and x=0.6 -- right.
} 
\end{figure}
\twocolumn

\end{document}